\documentclass{aa}
\usepackage{physics}
\usepackage{txfonts}
\usepackage{bbold}
\usepackage{graphicx}
\usepackage{placeins}
\graphicspath{{figures/}}
\usepackage[separate-uncertainty=true,multi-part-units=single]{siunitx}
\usepackage{wasysym}
\usepackage{hyperref}
\usepackage{xcolor}
\usepackage{gensymb}
\usepackage[T1]{fontenc}
\DeclareSIUnit\gauss{G}
\DeclareSIUnit{\mass}{M}
\DeclareSIUnit{\radius}{R}
\DeclareSIQualifier{\solar}{\astrosun}
\DeclareSIQualifier{\stellar}{_\star}
\usepackage{amsmath}
\usepackage{bm}

\newcommand\sfig[2]{\begin{figure} \resizebox{\hsize}{!}{\includegraphics{#1}}\caption{#2}\label{fig:#1} \end{figure}}
\newcommand\lfig[2]{\begin{figure*} \resizebox{\hsize}{!}{\includegraphics{#1}}\caption{#2}\label{fig:#1} \end{figure*}}

\makeatletter
\renewcommand*\aa@pageof{, page \thepage{} of \pageref*{LastPage}}
\makeatother

\title{Interior rotation modelling of the $\beta\,$Cep pulsator HD~192575 including multiplet asymmetries}
\newcommand\orc[1]{\href{https://orcid.org/#1}{\includegraphics[width=3mm]{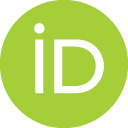}}}

\author{
  V.~Vanlaer\inst{\ref{kul}}\orc{0000-0003-4923-6199},
  D.~M.~Bowman\inst{\ref{NU},\ref{kul}}\orc{0000-0001-7402-3852},
  S.~Burssens\inst{\ref{kul}}\orc{0000-0002-1593-0863},
  S.~Bharati~Das\inst{\ref{ista},\ref{cfa}}\orc{0000-0003-0896-7972},
  L.~Bugnet\inst{\ref{ista}}\orc{0000-0003-0142-4000},
  S.~Mathis\inst{\ref{saclay}}\orc{0000-0001-9491-8012},
  C.~Aerts\inst{\ref{kul},\ref{imapp},\ref{mpiaheidelberg}}\orc{0000-0003-1822-7126}
}
\abstract{
  Rotation plays an important role in stellar evolution. However, the mechanisms behind the transport of angular momentum in stars at various stages of their evolution are not well understood. To improve our understanding of these processes, it is necessary to measure and validate the internal rotation profiles of stars across different stages of evolution and mass regimes. 
}
{
  Our aim is to constrain the internal rotation profile of the 12-M$_\odot$ $\beta$~Cep pulsator HD~192575 from the observed pulsational multiplets and the asymmetries of their component frequencies.
}
{
  We update the forward asteroseismic modelling of HD~192575 based on new TESS observations.
  We invert for the rotation profile from the symmetric part of the splittings, and compute the multiplet asymmetries due to the Coriolis force and stellar deformation treated perturbatively.
  We compare the computed asymmetries with the observed asymmetries.
}
{
  Our new forward asteroseismic modelling is in agreement with previous results, but with increased uncertainties, partially due to increased frequency precision, requiring us to relax certain constraints.
  Ambiguity in the mode identification is the main source of the uncertainty, which also affects the inferred rotation profiles.
  Almost all acceptable rotation profiles occur in the regime below 0.4~d$^{-1}$ and favour weak radial differential rotation, with a ratio of core and envelope rotation below two.
  We find that the quality of the match between the observed and theoretically predicted mode asymmetries is strongly dependent on the mode identification and the internal structure of the star.
}
{
  Our results offer the first detailed rotation inversion for a $\beta\,$Cep pulsator.
  They show that the rotation profile and the mode asymmetries provide a valuable tool to further constrain the evolutionary properties of HD~192575, and in particular the details of angular momentum transport in massive stars.
}


\date{Received ??? / Accepted ???}
\institute{Institute of Astronomy, KU Leuven, Celestijnenlaan 200D, 3001, Leuven, Belgium \label{kul}\\
  \email{vincent.vanlaer@kuleuven.be} \and
  School of Mathematics, Statistics and Physics, Newcastle University, Newcastle upon Tyne, NE1 7RU, UK \label{NU} \and
  Institute of Science and Technology Austria (ISTA), Am Campus 1, Klosterneuburg, Austria \label{ista} \and
  Center for Astrophysics | Harvard \& Smithsonian, 60 Garden Street, Cambridge, MA 02138, USA \label{cfa} \and
  Université Paris-Saclay, Université Paris Cité, CEA, CNRS, AIM, 91191, Gif-sur-Yvette, France \label{saclay}\and
  Department of Astrophysics, IMAPP, Radboud University Nijmegen, PO Box 9010, 6500 GL Nijmegen, The Netherlands \label{imapp} \and
  Max Planck Institute for Astronomy, K\"onigstuhl 17, 69117, Heidelberg, Germany \label{mpiaheidelberg}
}
\authorrunning{Vanlaer et al.}

\begin{document}
\maketitle

\section{Introduction}

The transport of angular momentum in stars with a convective core as measured through asteroseismology does not align with theoretical predictions for a significant number of stars \citep{eggenberger-angular-2012,marques-seismic-2013,goupil-seismic-2013,ceillier-understanding-2013,cantiello-angular-2014,fuller-angular-2014,belkacem-angular-2015,ouazzani-doradus-2019,moyano-asteroseismology-2022,salmon-backtracing-2022,betrisey-testing-2023} --- see \citet{aerts-angular-2019} for a review.
Various mechanisms such as magnetic fields \citep[e.g.][]{mestel-magnetic-1987,spruit-dynamo-2002,mathis-transport-2005,fuller-slowing-2019,takahashi-modeling-2021} and internal gravity waves \citep[e.g.][]{schatzman-transport-1993,zahn-angular-1997,talon-angular-2002,rogers-internal-2013,fuller-angular-2014,rogers-differential-2015} are proposed as possible solutions to this inconsistency.
In order to validate such proposals, it is necessary to measure the rotation profile of stars at various masses and stages of evolution.
However, the internal rotation profiles of only a handful of massive stars have been constrained from their oscillation frequencies \citep{bowman-asteroseismology-2020}.

The $\beta$\,Cep star HD~192575 discovered in TESS space photometry is a unique calibrator for interior rotation \citep{burssens-calibration-2023}.
While studies of $\beta\,$Cep stars go back more than a century \citep[see the reviews by][]{sterken-beta-1993,aerts-cep-2003,stankov-catalog-2005,bowman-asteroseismology-2020}, few such stars have been modelled asteroseismically and fewer still have a reliable measurement of their internal rotation rates \citep{aerts-asteroseismology-2003,pamyatnykh-asteroseismology-2004,dupret-asteroseismology-2004,briquet-asteroseismic-2007,dziembowski-two-2008,suarez-seismology-2009,salmon-backtracing-2022}.
HD~192575 was modelled asteroseismically by \citet{burssens-calibration-2023} based on the detection of rotationally split multiplets of low-order pressure (p) and gravity (g) modes in its frequency spectrum deduced from 352~d of cycle 2 TESS space photometry.
Three of these multiplets, two g-mode and one p-mode multiplet, were used in the subsequent forward asteroseismic modelling.
HD~192575 was found to be a differentially rotating star with a mass of $12.0^{+1.5}_{-1.5}$\SI{}{\mass\solar} approaching the end of its main-sequence life with central hydrogen mass fraction of $X_{\rm c} = 0.176^{+0.035}_{-0.040}$.

Internal rotation profiles of slowly rotating stars have been derived from the symmetric component of rotationally split multiplets \citep[e.g.][]{deheuvels-seismic-2012,deheuvels-seismic-2014,kurtz-asteroseismic-2014,dimauro-internal-2016,dimauro-rotational-2018,triana-internal-2015,triana-internal-2017,hatta-twodimensional-2019,hatta-bayesian-2022,aerts-asteroseismic-2024}, but moderate-to-fast rotation also causes asymmetrical splittings \citep{saio-rotational-1981,gough-effect-1990,soufi-effects-1998,suarez-effects-2006}.
Other effects also contribute to these asymmetries, such as the presence of a magnetic field \citep[e.g.][]{gough-effect-1990,mathis-probing-2021,bugnet-magnetic-2021,loi-topology-2021,bugnet-magnetic-2022,mathis-asymmetries-2023,bharatidas-unveiling-2024,guo-oscillation-2024,bhattacharya-detectability-2024}, which has been used to derive the internal magnetic field strengths of red giants \citep{li-magnetic-2022,deheuvels-strong-2023}.
Characterising the effect of rotation on the asymmetries, and validating the consistency of the observed asymmetries with these theoretical predictions is both useful to constrain the rotation profile, and to provide a starting point for further investigation into these additional contributions.
Thus far, only the asymmetries of two $\beta\,$Cep stars, $\nu\,$Eri and $\theta\,$Oph, have been analysed in detail \citep{pamyatnykh-asteroseismology-2004,briquet-asteroseismic-2007,suarez-seismology-2009}.

We aim to derive the internal rotation profile of HD~192575 from its observed multiplet splittings and investigate whether the theoretical asymmetries match the observed ones, given the new rotation profile.
First, we determine the internal rotation profile from the symmetric part of the splittings of HD~192575 using more recently available (i.e. cycles 2 and 4) TESS data (Sect.~\ref{sec:new-sectors}) via rotation inversion (Sect.~\ref{sec:inversion}).
Using the candidate rotation profiles, we compute rotational asymmetries for the modelled modes based on a perturbative approach in Sect.~\ref{sec:rot-asymmetries}.
These asymmetries are then compared to the observed asymmetries to check whether they are consistent.
Finally, we summarize and conclude in Sect.~\ref{sec:conclusions}.

\section{New asteroseismic modelling}
\label{sec:new-sectors}

\lfig{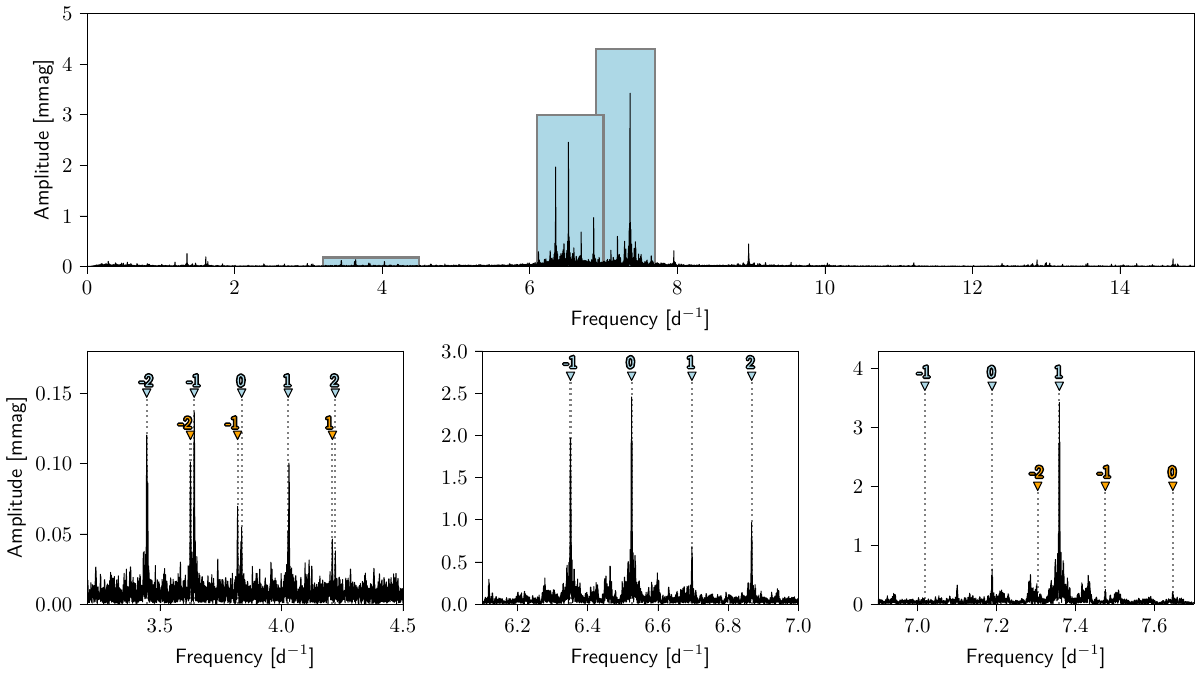}{Frequency spectrum of HD~192575. The top panel shows the full frequency spectrum, with the regions with fitted modes indicated by the light blue boxes. The bottom panels show zoom-ins of those boxes. From left to right, the multiplets are given the names quint1a (blue, first panel), quint1b (orange, first panel), quad1 (blue, second panel), trip1 (blue, third panel), and trip2 (orange, third panel).}

HD~192575 is an early-B dwarf that continues to be observed by the Transiting Exoplanet Survey Satellite in its Northern continuous viewing zone \citep[TESS; ][]{ricker-transiting-2014}.
It has a rich frequency spectrum with multiple g- and p-mode rotationally-split multiplets, with frequencies ranging from \SI{3.5}{\per\day} to \SI{15}{\per\day}.
One of the features of this frequency spectrum is the overlap of two multiplets in the g-mode dominated regime between \SI{3.5}{\per\day} and \SI{4.5}{\per\day} cycles per day (we refer to these as quint1a and quint1b, see Fig.~\ref{fig:LC+PG_panel_HD192575}).
\citet{burssens-calibration-2023} identified these as two $\ell = 2$ multiplets, and therefore they must be adjacent in radial order, and by necessity undergoing an avoided crossing with each other.
This was used by \citet{burssens-calibration-2023} to constrain the properties of HD~192575.

In this paper, we update this modelling as TESS has revisited HD~192575 during its extended mission, yielding new light curves and therefore revised oscillation frequencies and mode identifications.
The results of these new modelling efforts can be found in Table~\ref{table:parameter-ranges}.
We use the same grid of MESA \citep[Modules for Experiments in Stellar Astrophysics;][]{paxton-modules-2011,paxton-modules-2013,paxton-modules-2015,paxton-modules-2018,paxton-modules-2019,jermyn-modules-2023} and GYRE \citep{townsend-gyre-2013,townsend-angular-2018,goldstein-contour-2020} models and the same modelling techniques and software as \citet{burssens-calibration-2023}.
The only difference, as explained below, between our work and the work of \citet{burssens-calibration-2023} is the updated observed mode frequencies and the mode identification based on new TESS data.

At the time of the asteroseismic modelling of HD~192575 by \citet{burssens-calibration-2023}, only cycle 2 of TESS observations was available.
Meanwhile, TESS has revisited HD~192575 during its extended mission.
We combined all the available TESS data for this star until the end of 2023.
We extracted the 2-min cadence PDC-SAP light curves from MAST\footnote{\url{https://archive.stsci.edu/missions-and-data/tess}}, applied the same reduction and detrending routines as \citet{burssens-calibration-2023}, and reanalysed them using the iterative pre-whitening routines of \citet{bowman-systematic-2021}.
The pulsation mode frequencies extracted for the g-mode and p-mode multiplets based on the combined cycles 2 and 4 TESS light curves are listed in Appendix~\ref{sec:frequencies}.

Due to the extended time base of the combined cycles 2 and 4 TESS light curve, some pulsation mode frequencies, which previously were unresolved or statistically insignificant compared to the noise in the frequency spectrum, are now identified as significant.
This includes an additional mode in the quint1a multiplet, requiring a different assignment of the azimuthal orders compared to the identification by \citep{burssens-calibration-2023}.
The zonal mode of this multiplet is now set at the \SI{3.84}{\per\day} mode instead of the \SI{3.64}{\per\day} mode (see Fig.~\ref{fig:LC+PG_panel_HD192575}).

In their work, \citet{burssens-calibration-2023} limited the selection of valid models by enforcing an avoided crossing criterion, meaning that the frequency difference between the two identified g-mode multiplets should be similar to the observed frequency difference, allowing for observational uncertainties.
With the extended time base of the TESS observations, this observed difference has decreased by about a factor two.
Therefore, we re-apply the same forward asteroseismic modelling framework for HD~192575 with these stricter constraints.
We find that the radial orders of the modes increase by unity, that is the $\ell = 2$ g\textsubscript{2}, g\textsubscript{1}, and p\textsubscript{2} modes are now identified as g\textsubscript{1}, f, and p\textsubscript{3}, respectively.
Using the same constraints on effective temperature and luminosity derived by \citet{burssens-calibration-2023}, a spectroscopic selection criterion using 2-$\sigma$ confidence intervals in the Hertzsprung--Russell (HR) diagram, only 7 structure models in the grid of {\sc MESA} models provide appropriate matches of predicted frequencies that fit the TESS data.
In order to maintain a reasonably wide sample of forward models, we relax the spectroscopic confidence intervals to 3$\sigma$, with 42 {\sc MESA} structure models surviving this cut.
As was done by \citet{burssens-calibration-2023} we make a cut-off for low $X_{\rm c}$, keeping only models with $X_{\rm c} \geq 0.05$, to drop outliers before computing the confidence intervals.
These evolved models are separated in the HR diagram from the other models and have a much lower likelihood (about one to two orders of magnitude) than the best fitting models, but would skew the confidence intervals in Table~\ref{table:parameter-ranges} significantly.
Additionally, they have a lower probability of being observed, owing to the short-lived contraction phase near the terminal-age main sequence.
The resulting 2-$\sigma$ confidence intervals on the structure parameters, such as mass and age, are similar to the original modelling results of \citet{burssens-calibration-2023}.
Thus the results of the forward asteroseismic modelling in this work are comparable to those performed previously, meaning they are not so sensitive to the updated mode identification.
In the remainder of this paper we refer to this new set of best-fitting forward models as the ``new frequencies'' model set.

We now consider alternative explanations for the two close g-mode multiplets, as this is the most constraining factor in the forward asteroseismic modelling.
For example, one of the multiplets could be a higher angular degree multiplet (e.g. $\ell = 3$, $\ell = 4$) that purely by chance ended up in a similar frequency location, and with similar splittings.
Another possibility is that there is only one multiplet, but it is split by both rotation and a misaligned magnetic field \citep[e.g.][]{loi-topology-2021}.
We allow for these possibilities by fitting only one of the two g-mode multiplets.
Due to the proximity of these multiplets, the choice of which multiplet to keep, and which multiplet to discard, does not affect our use case.
We present and use the result for the forward asteroseismic modelling with the quint1a multiplet, ignoring the quint1b multiplet.
Furthermore, we attempt to fit three of the higher-frequency observed multiplets: the quadruplet already fitted by \citet{burssens-calibration-2023} (quad1), the triplet containing the dominant mode (trip1), and one more triplet (trip2).
These selections were made based on an initial fit of only quint1a and trip1, where we assume that quint1a is an $l=2$ mode and trip1 is an $l=1$ mode.

With the constraint from the avoided crossing between the g-mode multiplets removed, multiple sets of mode identifications are possible.
We constrain the mode degrees to $\ell = 1$ and $\ell = 2$, as none of the multiplets consist of more than five frequencies, and higher degree modes are less likely to be observed due to geometric cancellation effects.
Once the mode identifications of quint1a and trip1 are fixed, the other modes (quad1, trip2) are also fixed, as only the frequencies of a handful of radial orders fall in the region with the observed multiplets.
We find that quad1 and trip2 should both be $\ell = 2$ modes, with adjacent radial orders.

We selected six sets of mode identifications that fit the observed frequencies best.
All mode identifications used can be found in Table~\ref{table:mode-id}.
Confidence intervals are provided in Table~\ref{table:parameter-ranges} for the associated parameters.
It is worth noting that there are multiple candidates in the quad1 and trip2 multiplet for the zonal mode, and that different choices for the zonal mode do not yield the exact same forward modelling results.
However, the variation introduced by this freedom is smaller than the variation between the different mode identifications of the multiplets.
Hence, we only consider one choice for the zonal mode, given in Table~\ref{table:frequencies}.

Compared to the restricted modelling results, the central hydrogen fraction, $X_{\rm c}$, for these six model sets is less constrained, as we no longer rely on the constraining power of the avoided crossing (see Table~\ref{table:parameter-ranges}).
This also generally increases the confidence intervals of all the inferred parameters, such as the convective core mass.
The estimated model parameters are almost identical among the different sets of models.
The initial stellar mass ranges from $\SI{10.5}{\mass\solar}$ to $\SI{13.5}{\mass\solar}$ for all but one of the model sets, while the metallicity and the convective boundary mixing (CBM) parameter, $f_{\rm CBM}$, are effectively unconstrained, spanning the entire parameter space, which was also reported by \citet{burssens-calibration-2023} owing to inherent model parameter degeneracies.
It is important to note that the $2$-$\sigma$ confidence intervals quoted in Table~\ref{table:parameter-ranges} do not take into account the different correlation structures across the different model sets.
While we cannot compare the likelihood of Set 1 to 6 with the original and ``new frequencies'' model sets, as they use different observed quantities, it is possible to compare sets 1 to 6 amongst themselves.

Models close to the end of the main sequence are less likely to be observed from a evolutionary timescale standpoint.
The further along its main-sequence life the star is, the faster it burns the remaining hydrogen in its core. The main sequence is followed by a fast contraction phase, where a significant fraction of energy is generated by gravitational collapse whilst a small fraction of hydrogen remains in the core.
Hence, stars with lower $X_c$ should be observed less likely in nature.
This effect was not directly taken into account by \citet{burssens-calibration-2023} when computing the likelihood of a certain stellar model.
Therefore, our modelling also does not take this effect into account, as we use the same modelling strategy as \citet{burssens-calibration-2023} for reasons of proper comparisons among the results.
Figure~\ref{fig:xc_pdf_1} shows the probability density function of observing one of our evolutionary tracks as a function of $X_{\rm c}$.
In the range given by our models, the likelihood of finding the star changes approximately by a factor of two, which is another factor of two less likely compared to a star close to the zero-age main sequence.
While low $X_c$ are less likely, the difference is not so large as to completely discredit models at $X_c$ close to zero.
This effect of timescales would be most relevant for the selection of the best model for each of the model sets, especially for model sets 4 through 6.

From a population viewpoint, $\beta$~Cep stars with $X_c < 0.1$ could be considered outliers \citep[who used the same model grid as we do]{fritzewski-mode-2025}.
All of the model sets have solutions within this region, and it may be reasonable to consider such solutions as less likely.
Similar to the timescale effect from the previous paragraph, we do not include the population in the forward modelling to remain consistent with \citet{burssens-calibration-2023}.

{\renewcommand{\arraystretch}{1.3}
\begin{table*}
  \caption{Best model parameters and 2-$\sigma$ confidence intervals for the forward asteroseismic modelling of HD~192575 for the original modelling by \citet{burssens-calibration-2023}
  and our seven new modelling sets.}
  \label{table:parameter-ranges}
  \centering
  \begin{tabular}{l c c c c c c c c}
    \hline
    \hline
    Parameter & Original & New frequencies & Set 1 & Set 2 & Set 3 & Set 4 & Set 5 & Set 6\\
    \hline
    $Z_{\rm ini}$ [dex] & $0.012^{+0.004}_{-0.000}$ & $0.016^{+0.000}_{-0.004}$ & $0.014^{+0.002}_{-0.002}$ & $0.012^{+0.004}_{-0.000}$ & $0.016^{+0.000}_{-0.004}$ & $0.014^{+0.002}_{-0.002}$ & $0.014^{+0.002}_{-0.002}$ & $0.012^{+0.004}_{-0.000}$ \\
$M_{\rm ini}$ [M$_\odot$] & $12.0^{+1.500}_{-1.500}$ & $12.0^{+2.000}_{-0.000}$ & $13.0^{+0.500}_{-2.500}$ & $11.5^{+2.000}_{-1.000}$ & $11.0^{+2.500}_{-0.500}$ & $12.5^{+1.000}_{-2.000}$ & $12.0^{+1.500}_{-1.500}$ & $12.0^{+1.500}_{-1.500}$ \\
$f_{\rm CBM}$ & $0.030^{+0.005}_{-0.025}$ & $0.035^{+0.000}_{-0.020}$ & $0.035^{+0.000}_{-0.030}$ & $0.035^{+0.000}_{-0.030}$ & $0.025^{+0.010}_{-0.020}$ & $0.010^{+0.025}_{-0.005}$ & $0.010^{+0.025}_{-0.005}$ & $0.015^{+0.020}_{-0.010}$ \\
$X_{\rm c}$ & $0.176^{+0.035}_{-0.040}$ & $0.156^{+0.000}_{-0.035}$ & $0.236^{+0.025}_{-0.220}$ & $0.136^{+0.175}_{-0.120}$ & $0.251^{+0.160}_{-0.235}$ & $0.016^{+0.285}_{-0.000}$ & $0.026^{+0.225}_{-0.010}$ & $0.016^{+0.395}_{-0.000}$ \\
\hline
$M_{\rm cc}$ [M$_\odot$]  & $2.87^{+0.48}_{-0.81}$ & $2.92^{+0.42}_{-0.06}$ & $3.56^{+0.07}_{-1.96}$ & $2.68^{+0.98}_{-1.07}$ & $2.56^{+1.12}_{-0.96}$ & $2.14^{+1.43}_{-0.40}$ & $2.03^{+1.55}_{-0.42}$ & $2.10^{+1.35}_{-0.50}$ \\
$R_{\rm cc}$ [R$_\odot$]  & $0.91^{+0.11}_{-0.15}$ & $0.94^{+0.07}_{-0.01}$ & $1.05^{+0.02}_{-0.44}$ & $0.87^{+0.20}_{-0.27}$ & $0.87^{+0.20}_{-0.27}$ & $0.71^{+0.34}_{-0.07}$ & $0.71^{+0.35}_{-0.11}$ & $0.70^{+0.33}_{-0.10}$ \\
$R_{\rm *}$ [R$_\odot$]  & $9.1^{+0.8}_{-1.7}$ & $10.2^{+0.7}_{-0.2}$ & $8.8^{+2.4}_{-1.6}$ & $10.2^{+1.1}_{-3.4}$ & $7.4^{+3.6}_{-1.4}$ & $10.2^{+1.0}_{-3.4}$ & $9.9^{+1.3}_{-2.7}$ & $10.4^{+0.8}_{-4.4}$ \\
    \hline
  \end{tabular}
  \tablefoot{The ``original'' and ``new frequencies'' refer to the modelling sets with the avoided crossing criterion, while ``Set 1'' to ``Set 6'' only consider one of the merged g-mode multiplets. Both fitted parameters ($Z_{\rm ini}$, $M_{\rm ini}$, $f_{\rm CBM}$, and $X_{\rm c}$) and inferred values (convective core mass $M_{\rm cc}$ and radius $R_{\rm cc}$, and stellar radius $R_*$) are given.}
\end{table*}
}

\begin{table*}
  \caption{Mode identification of the multiplets used for the different model sets, given as $(\ell, n_{pg})$, where a dash indicates a multiplet was not included for that specific model set.}
  \label{table:mode-id}
  \centering
  \begin{tabular}{l c c c c c c c c}
    \hline
    \hline
    Multiplet & Original & New frequencies & Set 1 & Set 2 & Set 3 & Set 4 & Set 5 & Set 6\\
    \hline
    quint1a & $(2, -2)$ & $(2, -1)$ & $(2, -1)$ & $(2, -1)$ & $(2, -2)$ & $(2, -2)$ & $(2, -2)$ & $(2, -2)$ \\
    quint1b & $(2, -1)$ & $(2, +0)$ & - & - & - & - & - & - \\
    quad1   & $(2, +2)$ & $(2, +2)$ & $(2,  +2)$ & $(2, +3)$ & $(2, +0)$ & $(2, +1)$ & $(2, +2)$ & $(2, +3)$ \\
    trip1   & - & - & $(1, +4)$ & $(1, +5)$ & $(1, +2)$ & $(1, +3)$ & $(1, +4)$ & $(1, +5)$ \\
    trip2   & - & - & $(2, +3)$ & $(2, +4)$ & $(2, +1)$ & $(2, +2)$ & $(2, +3)$ & $(2, +4)$ \\
    \hline
  \end{tabular}
\end{table*}

\sfig{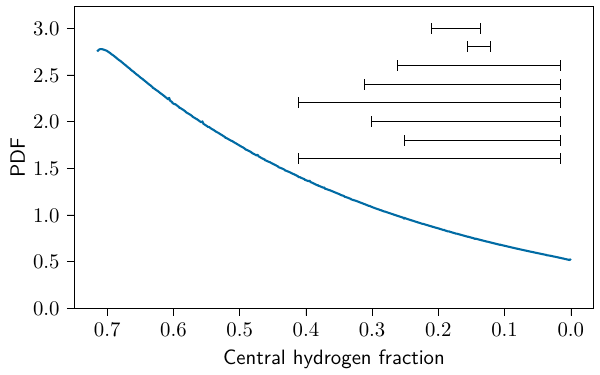}{Probability density function of observing one of the models at a certain central hydrogen fraction. The intervals (black) show the uncertainty regions for the central hydrogen fraction for each of the model sets (see Table~\ref{table:parameter-ranges}). From top to bottom, the intervals show the ``original'', ``new frequencies'', and sets 1 through 6 solutions. }

\section{Rotation inversion}
\label{sec:inversion}

\subsection{Setup}

In this section, we infer the rotation profile of HD~192575 from the observed multiplets.
We use the symmetric part (the average of all splittings) of these multiplets as input for the rotation inversion, as the asymmetric components depend on the rotation frequency only at the second or higher order.
The uncertainties on these symmetric components include both the uncertainty on the observed mode frequencies and the range of splittings for each multiplet, taken as a 1-$\sigma$ uncertainty.
This second effect is dominant, as the observational uncertainties are on the order of $10^{-6}~\SI{}{\per\day}$, while the asymmetries are on the order of $10^{-3}~\SI{}{\per\day}$.
For each of the model sets discussed in Sect.~\ref{sec:new-sectors}, we consider the 100 best-fitting models, except for the new frequencies model set, as it only contains 42 models within the 3-$\sigma$ confidence intervals for effective temperature and luminosity.

As in \citet{burssens-calibration-2023}, we use the Mahalanobis distance \citep{mahalanobis-generalized-1936} with an additional term for the theoretical variance of the stellar models 
as merit function \citep{aerts-forward-2018}.
Some models appear in multiple model sets, but the inferred rotation profiles are distinct, as the mode identification for each model set is different.

We perform a regularized least-squares (RLS) fit \citep[e.g.][]{christensen-dalsgaard-comparison-1990} for each of the models to derive the internal rotation profile of HD~192575.
RLS is a linear least-squares fit, with an additional penalty term for non-regular profiles.
This penalty term is multiplied by a regularization parameter to balance the importance of the observations and the level of regularity.
We fit the same frequencies with the same associated mode identifications as those used in the forward asteroseismic modelling in Section~\ref{sec:new-sectors}.
The basis functions are degree two b-splines with 14 knots, which are equally spaced from the centre of the star to the surface, except for the endpoints (i.e. the centre of the star and the surface), where three knots are placed at the same location.
This is to ensure that the resulting spline basis functions are localized and do not gain large tails near the edges of the interval.
This gives us a total of eleven control points, which are free parameters in the fit.
The rotation profile is regularized by the norm of its first derivative, which is integrated over the entire star.
We vary the regularization parameter such that we adequately cover the range around a $\chi_{\rm obs}^2$ error of unity, therefore allowing for potentially under- and over-fitting.
The $\chi_{\rm obs}^2$ error includes only the observational errors, and not the contribution from the regularization in the optimization:
\begin{equation}
    \chi_{\rm obs}^2 = \sum_i \frac{(\Delta^{\rm obs}_i - \Delta^{\rm model}_i)^2}{\sigma^2_i}\quad,
\end{equation}
where $\Delta^{\rm obs}_i$ and $\Delta^{\rm model}_i$ are the observed and modelled multiplet splittings, and $\sigma_i$ are the uncertainties on these splittings, which include the range spanned by the asymmetries.

In order to select the final rotation profiles, we consider two criteria: (i) the L-curve criterion \citep{hansen-lcurve-2001}; and (ii) checking that the rotation profiles can actually reproduce the observed splittings.
The L-curve is a log-log plot comparing the goodness of fit, given by $\chi_{\rm obs}^2$, and how much the profile deviates from a ``regular'' profile, as defined by the regularization condition.
It can help determine optimal values for the regularization parameter by looking for parts of the diagram that have a similar shape as the letter L.
The value of the regularization parameter that results in the point in the corner would then be the optimal value \citep{hansen-lcurve-2001}.
For this second criterion, we take the rotation profile with the largest regularization parameter that still has the splittings within the observed uncertainty region.
We use this method instead of more commonly used point measures such as the optimally localized averages family of methods \citep[e.g.][]{pijpers-sola-1994} since we need a full rotation profile from the centre to the surface of the star to be able to compute mode asymmetries in the next Section.

Apart from constructing rotation profiles to understand the asymmetry of the multiplets, we also compare with the results from forward asteroseismic modelling performed by \citet{burssens-calibration-2023} and with the rotation profiles computed by \citet{mombarg-first-2023} using the 2D stellar structure and evolution code ESTER \citep{espinosalara-selfconsistent-2013, mombarg-first-2023}.
\citet{burssens-calibration-2023} considered only monotonic rotation profiles, consisting of a core rotation rate, a surface rotation rate, with a transition region between the two.
They considered multiple permutations for the location and extent of this transition region: (i) the region where CBM takes place; (ii) the region where a gradient exists in the mean-molecular-weight ($\mu$) left behind by the receding core; and (iii) the entire envelope.
In each case, the core was found to be rotating faster than the surface by at least a factor 1.2 and up to factor of 12.8, including observational uncertainties and uncertainties on the forward asteroseismic modelling.
The ESTER modelling includes the cumulative effect of rotational mixing due to meridional circulation along the main sequence, and allows one to obtain rotation profiles that are consistent with the evolutionary history of the star.
Given the same initial mass, metallicity and central hydrogen fraction, \citet{mombarg-first-2023} predict a monotonous rotation profile, decreasing from the core to the surface.
We take the model that is at the same central hydrogen fraction as HD~192575 as determined by the original modelling by \citet{burssens-calibration-2023}, and compute the rotation profile averaged over the azimuthal direction from this ESTER structure model.

\subsection{Rotation profiles}

We first discuss the resulting rotation profiles in terms of the L-curve criterion.
L-curves of the best forward model of each model set, can be found in Fig.~\ref{fig:l_curves_2}.
In our case, this refers to the integral of the norm of the derivative of the rotation profile.
For our models, the L-curve does not always have the expected L shape (e.g. see the original and Set 3 curves in Fig.~\ref{fig:l_curves_2}).
Additionally, for the models where the L-curve does prefer a value, $\chi_{\rm obs}^2$ is large, on the order of $10^3$.
While in principle this might be an optimal point for the criteria behind the L-curve, the differences between the observed splittings and the fit is larger than the difference in splitting between the g-mode and p-mode multiplets.
This means that the rotation profile is mostly dominated by the regularization, and not the actual observed splittings.
Hence, we do not use the L-curve in the selection of the profiles, and only use our second criterion based on $\chi_{\rm obs}^2$ itself.

The rotation profiles of the best fitting forward model (see Table~\ref{table:parameter-ranges}) of each model set with their observational uncertainties are shown in Fig.~\ref{fig:spline_rls_rotation_profiles_best}.
The full set of rotation profiles can be found in Appendix~\ref{sec:all-rotation-profiles}.
The different models show quite different behaviour in the rotation profiles, with most showing oscillating profiles, while only a few show a monotonic profile.
This difference in behaviour is larger than the 2-$\sigma$ uncertainty regions, based on the observational uncertainties on the oscillation frequencies, including the lack of certainty regarding the identity of the symmetric component of the splittings.
As these uncertainties include the range of splittings observed within one multiplet, they should not be considered to be statistical uncertainties but rather as giving a qualitative indication of frequency uncertainties.
Overall, the difference between the models in the specific model sets and the observational uncertainties show smaller differences than between the different model sets.
This indicates that the forward modelling, and specifically the mode identification, is the main source of uncertainty on the inverted profiles, making it challenging to draw conclusions about the rotation profiles.
We can also see this in the rotation kernels (see Fig.~\ref{fig:kernels_1}), which change appreciably between different models.
The rotation kernels define the relation between the observed frequencies and the rotation profile in the star.
Therefore, if the rotation kernels change significantly, either because the mode identification is different or the internal structure of the model is changed, so too does the rotation profile.

We find from considering the rotation profiles for all the forward models, that the additional modes considered for model sets 1 through 6 do exclude a monotonic decreasing profile.
This is not the case for the ``original'' and ``new frequencies'' modelling sets, where some models can still be described adequately with monotonic decreasing profiles (see Figs.~\ref{fig:rotation-profiles/spline_rls_rotation_profiles_original} and ~\ref{fig:rotation-profiles/spline_rls_rotation_profiles_increased_npg}).

Figure~\ref{fig:spline_rls_rotation_profiles_center_to_surface} shows the centre to near-surface rotation ratios for all the models in our model sets.
The rotation profiles were chosen in the same way as in the previous paragraph.
We find that large ratios of the core and surface rotation rates (i.e. $>5$) are not required by the observations, and that most profiles have a similar centre and near-surface rotation rate.
This does not mean that the rotation profiles themselves are near rigid, as can be seen in Fig.~\ref{fig:spline_rls_rotation_profiles_best}.
We note that the actual surface rotation rate for some of the models in model Set 5 are close to zero, which would make the core-to-surface rotation rate ill-defined.
Hence we take the rotation rate at 0.8R$_\star$ in the envelope as an approximation for the surface rotation rate.

The projected surface rotational velocity was obtained from spectroscopic observations by \citet{burssens-calibration-2023} and found to be $27^{+6}_{-8}$~km\,s$^{-1}$.
Given the possible range in stellar radii obtained from the forward modelling, the lower limit on the rotation frequency would be roughly 0.035~d$^{-1}$, which is in line with the results we obtain here, as most of the surface rotation rates are in the range 0.1~d$^{-1}$ to 0.3~d$^{-1}$.
Only some of the models in model Set~5 would be inconsistent with this lower limit (see Fig.~\ref{fig:rotation-profiles/spline_rls_rotation_profiles_final_set_5}).

We experimented with changing the parameters for the RLS: changing the number of spline knots to add and their position, and switching the first derivative regularization to a second derivative regularization.
We found that the effects of changing the knots are negligible compared to the differences between the model structures.
We tested whether placing knots non-uniformly to better match the sensitivity of the kernels mattered, but again found no significant changes.
However, changing the regularization function did affect the results, leading to rotation profiles with a much larger amplitude, although the locations of the peaks and valleys are similar to those of the profiles from the first derivative regularization.
Therefore, a regularization based on the second derivative yields unphysical results.

Finally, we compare our rotation profiles to the rotation profile obtained by \citet{mombarg-first-2023} using the two-dimensional code ESTER, which is shown as the red dash-dotted line in Fig.~\ref{fig:spline_rls_rotation_profiles_best}.
As the model was calibrated to the rotation rate at the surface found by \citet{burssens-calibration-2023}, we find that the value for the surface rotation rate matches with our results.
The results from \citet{mombarg-first-2023} predict a monotonic increase from a rotation rate of \SI{0.2}{\per\day} at the surface to \SI{0.4}{\per\day} at the centre of the star
and are in good agreement with some of the profiles in our solution sets.
However, most of our 1D solutions are not monotonically increasing.
While ESTER includes important elements of modelling the rotation profile in a star, such as meridional circulation, in a self-consistent way, there are still aspects in terms of angular momentum transport that are not included, such as internal gravity waves \citep[e.g.][]{talon-hydrodynamical-2005, rogers-differential-2015} or magnetic fields \citep{mestel-magnetic-1987, spruit-dynamo-2002, fuller-slowing-2019}, which are relevant for stellar evolution \citep{aerts-angular-2019}.
Furthermore, one-dimensional evolution codes such as MESA may include other important descriptions of physical processes, notably CBM and other forms of mixing, that ESTER models do not include in the implementation used by \citet{mombarg-first-2023}.
While these results are too uncertain at this point to speculate on which transport mechanism needs to be considered or tweaked, this discrepancy still indicates that modifications to ESTER and/or aspects of the asteroseismic modelling need to be made, as subjects of future work.

\sfig{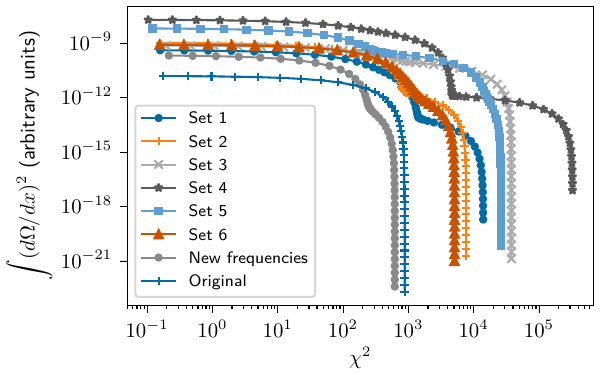}{L-curve for the best forward model (see Table~\ref{table:parameter-ranges}) from each model set, with the regularization parameter ranging from $10^3$ to $10^{10}$. The x-axis is the goodness-of-fit, given by the $\chi_{\rm obs}^2$. The y-axis shows the RLS penalty term, which is the integral of the square of the derivative of the profile. The optimal value for the regularization according to the L-curve criterion should be in the hook of the L shapes in the profiles.}
\sfig{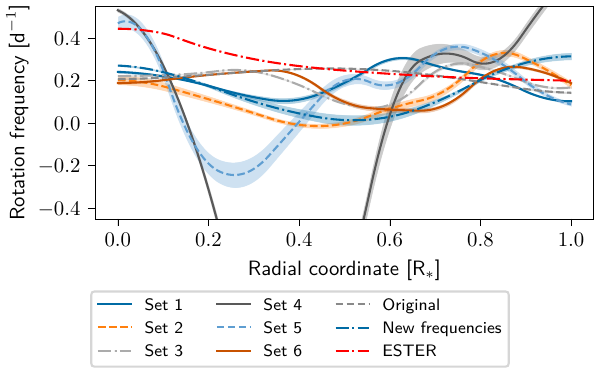}{Rotation profiles for the best model of each model set, with a regularization parameter set to the maximum possible value such that the profiles still reproduce the observed splitting (see main text for details). The bands around the profiles show the propagated 2-$\sigma$ uncertainties, including the lack of certainty regarding the identity of the symmetric component of the splittings. The red dash-dotted line shows the rotation profile obtained by \citet{mombarg-first-2023} using the ESTER code. The rotation profile of the best asteroseismic model of Set~4 ranges from \SI{-1.2}{\per\day} to \SI{0.7}{\per\day}.}
\sfig{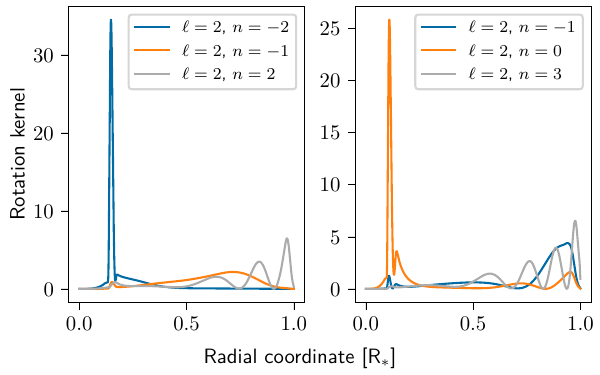}{Rotation kernels used in the rotation inversions for two of the forward models. The left panel shows the best model of the original model set, while the right panel shows the best model for the ``new frequencies'' model set. This plot shows how different the kernels can be between the different forward models. When matched to the same observations, the output from the rotation inversion can vary significantly between the models.}
\sfig{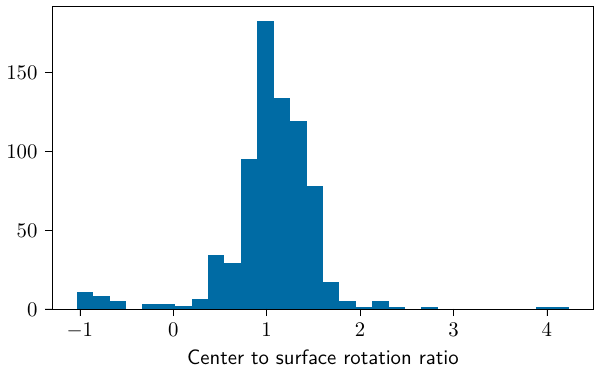}{Ratio of the rotation frequency in the centre of the star and at a radius of \SI{0.8}{\radius\stellar}. This choice of surface point was made to prevent ``pathological'' values, as for some of the model sets the profiles tend to values around zero for the surface rotation rate.}

\section{Rotational asymmetries of HD~192575}
\label{sec:rot-asymmetries}
\subsection{Higher-order effects of rotation}

Rotation has at least two main effects on the pulsation frequencies of a star.
It changes the evolution of a star through various transport processes and it causes a non-spherical deformation of the structure of the star, making it oblate  \citep[e.g.][]{maeder-physics-2009}.
Secondly, the linearized pulsation equations gain linear and quadratic terms in the rotation frequency due to the Coriolis acceleration, the centrifugal acceleration, and the rotational shear.
In combination with the structure changes from the star not being spherically symmetric, the solutions to the modified oscillation equations are no longer separable in $r$ and $\theta$.
This means that one either needs to make simplifying approximations, as we do in this paper, or use two-dimensional oscillation computations \citep{reese-acoustic-2006,ouazzani-pulsations-2012}.
So far, modelling of asymmetric splittings in $\beta\,$Cep pulsators have been limited, with the studies by \citet{pamyatnykh-asteroseismology-2004}, \cite{briquet-asteroseismic-2007}, \citet{suarez-seismology-2009}, and \citet{guo-oscillation-2024} as notable exceptions.
Here we go beyond their treatments in our modelling of HD\,192575's asymmetric low-order mode splittings, building upon the extensive modelling by \citet{burssens-calibration-2023} and the high-precision TESS-based observations.

Given the moderate rotation rate ($\sim$16\% of the Keplerian critical rotation rate) of HD~192575, we use a perturbative method to obtain the oscillation frequencies for a radially differential rotation profile, based on the formalisms of \citet{saio-rotational-1981} and \citet{lee-pulsational-1995}.
The oscillation equations we consider here contain the Coriolis force term, the structure deformation of the stellar model due to the rotation, the coupling between spherical harmonics with different degrees, and both the poloidal and toroidal components of the vector spherical harmonics.
The deformation of the stellar structure is computed using the Chandrasekhar-Milne expansion \citep[][]{chandrasekhar-equilibrium-1933} up to the second order $P_2$ contribution.
This assumes a constant rotation rate, which is not what we found for this star.
However, as the deformation is caused by the balance of the gravitational acceleration and the centrifugal acceleration, it is mostly felt in the outer layers of the star.
We therefore approximate the deformation using the rotation rate near the surface.
To compute other terms in the oscillation equation, we use the actual differential rotation profile, notably to evaluate the Coriolis force term.

We initially compute solutions of simplified oscillation equations consisting of the poloidal components of a single spherical harmonic.
This simplified system consists of the non-rotating terms and the Coriolis acceleration.
Since this is still a one-dimensional system, these equations can be solved non-perturbatively.
The solution for the poloidal component is valid up to first order in the rotation frequency. 
The Coriolis force also couples with the toroidal components of the $\ell \pm 1$ spherical harmonics.
We compute these components using analytical expressions \citep[e.g. Eqns.~(32) and (33) of][]{saio-rotational-1981}, which are also valid up to first order in the rotation frequency.
The combined poloidal and toroidal components are similar to the zeroth order system considered by \citet{soufi-effects-1998}.
This does not yet include the deformation of the stellar model by rotation and the feedback of the toroidal components of adjacent spherical degrees through the Coriolis force, both of which are second order in the rotation frequency.

We then approximate the eigenfunction of each oscillation mode for the full oscillation equations as a linear combination of eigenfunctions initially obtained from the simplified oscillation equations:
\begin{equation}
  \bm{\xi}^{\rm rot}_{k} = \sum_{k'}a_{kk'} \bm{\xi}_{k'}\,.
\end{equation}
Due to the azimuthal symmetry of the rotation, this linear combination only consists of oscillation modes with the same azimuthal order.
While in principle the sum over $k'$ is a sum over all radial orders and degrees, this is not possible in practice, and some truncation is needed.
Substituting this linear expansion in the full oscillation equations yields a quadratic eigenvalue problem for coefficients $a_{kk'}$ and the oscillation frequency $\omega$ as the eigenvalue.
More details on how such an eigenvalue problem can be solved can be found in Appendix~\ref{sec:quadratic-eigenvalue-problem}.
The physical terms included in this perturbation are the stellar deformation up to the $P_2$ Legendre polynomial, and the coupling with the toroidal modes through the Coriolis force.
This method is fundamentally the same as the methods described by \citet{saio-rotational-1981}, \citet{lee-pulsational-1995}, and \citet{soufi-effects-1998}, without the need to consider which modes should be included in computing near-degeneracy effects.

The perturbation method was validated with the two-dimensional oscillation code TOP \citep[][]{reese-acoustic-2006}, and shown to be sufficiently accurate to model the measured mode asymmetries of HD\,192575.
Detailed comparisons between our method and 2D computations with ESTER/TOP will be described elsewhere (Mombarg et al., in prep.).

\subsection{Observed asymmetries}
\label{sec:observed-asymmetries}

We define the asymmetries of a multiplet in the same way as \citet{burssens-calibration-2023}, namely as the difference between two adjacent splittings, such that:
\begin{equation} \label{eq: A_definition}
  \mathcal{A}_i = (f_{m = i} - f_{m = i-1}) - (f_{m = i-1} - f_{m = i-2})\,.
\end{equation}
For the quadrupole multiplets we therefore have $\mathcal{A}_0, \mathcal{A}_1$, and $\mathcal{A}_2$, which do not need to have the same value.

\sfig{rotation-asymmetries/spline_rls_o_c_trip1_a1}{Distribution of the theoretical asymmetries for the $\ell = 1$ multiplet. Each panel is a different set, indicated in the top left corner of the panel. The orange vertical line indicates the observed asymmetry. The radial orders associated with these asymmetries can be found in Table~\ref{table:mode-id} under the ``trip1'' multiplet. }

\sfig{rotation-asymmetries/spline_rls_o_c_quad1_a1}{Distribution of the theoretical $\mathcal{A}_1$ asymmetries for the ``quad1'' $\ell = 2$ multiplet. Each panel is a different set, indicated in the top left corner of the panel. The orange vertical line indicates the observed asymmetry. Since the multiplet is not complete, the $\mathcal{A}_1$ asymmetry has two observed values, which can be seen in the left panels as the two separate lines. The scale of the x axis in the right panel is too large to show the two values. The radial orders associated with these asymmetries can be found in Table~\ref{table:mode-id} under the ``quad1'' multiplet. }

In the case of model sets 1 through 6, the single dipole mode multiplet always has a negative asymmetry (see Fig.~\ref{fig:rotation-asymmetries/spline_rls_o_c_trip1_a1}).
The bulk of the models show an asymmetry on the order of $-0.1$~d$^{-1}$, which means that the distance between the retrograde mode and the zonal mode is about twice the distance between the zonal mode and the prograde mode.
The exact value varies significantly (up to about $-0.6$~d$^{-1}$), depending on the rotation frequency at the surface.
The observed asymmetry on the other hand, is on the order of 0.001~d$^{-1}$, more than two orders of magnitude smaller and positive.
Some of the models do get close to the observed values.
These correspond to models from model sets 3 and 5 with low surface rotation rates (see Figs.~\ref{fig:rotation-profiles/spline_rls_rotation_profiles_final_set_3} and \ref{fig:rotation-profiles/spline_rls_rotation_profiles_final_set_5}).
Out of the 600 models in model sets 1 through 6, 16 are within a value of 0.01~d$^{-1}$ from the observed value.

For the quadrupole modes, we find the asymmetries to be smaller by an order of magnitude for the same stellar models.
This can be seen in Fig.~\ref{fig:rotation-asymmetries/spline_rls_o_c_quad1_a1}, which shows the $\mathcal{A}_1$ asymmetry for the ``quad1'' multiplet.
Due to this multiplet only showing four of the five peaks needed for a full $\ell = 2$ multiplet, the $\mathcal{A}_1$ asymmetry has two possible observed values associated with it.
Only one of the values needs to fit the theoretical values, although both of them in the case of model sets 1, 3, and 5.
While we only show model sets 1 through 6, the results are similar for the original and new frequencies model sets.
The original model set is able to fit the data similarly well to sets 1, 3, and 5, while the new frequencies model set behaves similar to the other sets.
Given the two options for the $\mathcal{A}_1$ asymmetry, we either need to fit the $\mathcal{A}_0$ asymmetry or the $\mathcal{A}_2$ asymmetry.
We find that this is indeed feasible, and therefore it is possible to reproduce the asymmetries for this multiplet.
Of the remaining multiplets, we can reproduce the order of magnitude of the asymmetries of the ``quint1a'' multiplet with some of the model sets, as the theoretical g-mode asymmetries are smaller in this regime than the p-mode asymmetries.
The ``trip2'' multiplet is harder to reproduce, given its low asymmetry, with only set 3 being able to reproduce ``trip2''.
We provide supporting figures for all asymmetries not shown here in Appendix~\ref{sec:all-asymmetries}.

The asymmetries are strongly dependent on the individual model and its rotation profile.
Compared to the observational uncertainties, which are at least one order of magnitude smaller than the observed asymmetries, these asymmetries span a wide range of values. We have indicated which rotation profiles have corresponding asymmetries that are within $[0, 2\mathcal{A}]$ for all components of the observed ``quad1'' and ``quint1a'' asymmetries in Figs.~\ref{fig:rotation-profiles/spline_rls_rotation_profiles_original} through \ref{fig:rotation-profiles/spline_rls_rotation_profiles_final_set_6}.

\subsection{Discussion}

Our results show a variation in both the rotation profiles and therefore the corresponding theoretical asymmetries.
While these rotation profiles all provide a reasonable match with the symmetric parts of the observed rotationally split multiples, this is not the case for the ``trip1'' and ``trip2'' multiplets, and for a subset of the models for the ``quint1a'' and ``quad2'' multiplets.
Hence, the observed asymmetries could be used to constrain both the rotation profile, and the selection of the best forward model.
In order to fully benefit from this information, the forward modelling procedure and subsequent fitting and computational steps should be combined into one overarching modelling procedure.
It is encouraging, however, that our method finds asymmetries that are in agreement with the observations for two of the four multiplets, given that these were not considered at all during the stellar model selection process described in Sect.~\ref{sec:new-sectors}.

While improving the fitting procedures will lead to a more self-consistent analysis, there is still a discrepancy between the possible models and the observations for the ``trip1'' and ``trip2'' multiplets.
Several potential explanations for this discrepancy between the theoretical and observed asymmetries exist.
First and foremost, the 1D stellar models may not capture all the relevant physics needed to fully explain such detailed properties of the oscillation frequencies.
Second, the impact of the chosen chemical mixture for the models may be sub-optimal choices, impacting the mode cavities and kernel computations.
Third, a potential magnetic field, depending on its obliquity angle with the rotation axis, could counteract the rotational asymmetries (e.g. \citealt{loi-topology-2021}).
One of the alternative explanations of the double ``quint1a'' and ``quint1b'' multiplets in HD~192575 is the presence of such an internal magnetic field.
Finally, non-linear mode coupling may bring the asymmetries of the modes closer to zero \citep{buchler-dynamic-1995, buchler-role-1997}.
Our results are therefore suitable to be used in the future, with the aim to measure and constrain these properties in more detail.

\section{Conclusions}
\label{sec:conclusions}

In this paper, we performed an updated asteroseismic modelling of the $\beta$~Cep pulsator HD~192575 including new TESS mission data, fitted rotation profiles, and compared theoretical multiplet asymmetries with the observed asymmetries.
Our asteroseismic modelling results rely on more and more precise observables than those used by \citet{burssens-calibration-2023}. 
Keeping this in mind, our results agree well with the earlier ones achieved by \citet{burssens-calibration-2023}.
Uncertainty in the mode identification of the observed pulsation multiplets leads to a larger uncertainty on the age, as well as on the mass and radius of the convective core of HD~192575. While the initial mass, initial metallicity, and CBM have similar uncertainties due to degeneracies among these parameters, the uncertainty on the central hydrogen fraction has increased from $0.075$ to about 0.3, and the uncertainty on the inferred mass and radius of the convective core has increased by a factor two.

For a selection of 742 forward models within the resultant 2-$\sigma$ confidence intervals of the best solution, we obtained rotation profiles using RLS inversions.
The rotation profiles show a variety of behaviours, some having differential rotation on the order of 20\%, to profiles yielding counter-rotating regions.
While this may seem like a wide variety of option, we do stress that the rotation frequency within the star lies below \SI{0.4}{\per\day} for almost all profiles, in agreement with \citet{burssens-calibration-2023} and offering a remarkable calibration for theories of transport processes in massive stars.
The main uncertainty on the rotation profiles originates from the large differences among the 742 models owing to the mode identification and theoretical uncertainties in (1D) stellar evolution theory.
While firm conclusions about the properties of the rotation profile are challenging given these theoretical uncertainties, almost all forward models favour weak differential rotation, and the resultant core-to-surface rotation ratio is at most a factor of two for almost all models.

We computed theoretical asymmetries from the rotation profiles and compared these asymmetries with the observed asymmetries.
These computations include the effect of the Coriolis force and stellar deformation.
For two of the four multiplets, we find agreement between our model grid and the observed asymmetries, while for the other two multiplets the theoretical asymmetries are larger than the observed asymmetries.
We obtain different variations in computed asymmetries, depending on the mode identification and individual stellar model.

Our results show that the rotation profile and the associated asymmetries provide a valuable tool to further constrain the evolutionary properties of HD~192575, and in particular the details of angular momentum transport in massive pulsating stars.
To make maximal use of this information, we suggest that the future modelling endeavours of this star combine the forward modelling, rotation inversion, and computations of the asymmetries in a self-consistent modelling framework.
Such a framework currently does not exist but could then pave the way towards modelling more aspects of the evolution of the star, such as its internal magnetic field properties. 

\begin{acknowledgements}
  The authors appreciated the critical comments from the referee, which encouraged V.V. to embark upon a new code development sprint.
  V.V. gratefully acknowledges support from the Research Foundation Flanders (FWO) under grant agreement N°1156923N (PhD Fellowship) and N°K233724N (Travel grant).
  D.M.B. gratefully acknowledges support from the Research Foundation Flanders (FWO; grant number: 1286521N), and UK Research and Innovation (UKRI) in the form of a Frontier Research grant under the UK government's ERC Horizon Europe funding guarantee (SYMPHONY; grant number: EP/Y031059/1), and a Royal Society University Research Fellowship (URF; grant number: URF{\textbackslash}R1{\textbackslash}231631).
  S.B.D. acknowledges funding from the European Union’s Horizon 2020 research and innovation programme under the Marie Skłodowska-Curie grant agreement No 101034413.
  L.B. gratefully acknowledges support from the European Research Council (ERC) under the Horizon Europe programme (Calcifer; Starting Grant agreement N$^\circ$101165631).
  S.M. acknowledges support from the PLATO CNES grant at CEA/DAp.
  C.A. acknowledges financial support from the Research Foundation Flanders (FWO) under grant K802922N (Sabbatical leave); she is grateful for the kind hospitality offered by CEA/Saclay during her sabbatical work visits in the spring of 2023.
  The research leading to these results has received funding from the European Research Council (ERC) under the Horizon Europe programme (Synergy Grant agreement N$^\circ$101071505: 4D-STAR).  While funded by the European Union, views and opinions expressed are however those of the author(s) only and do not necessarily reflect those of the European Union or the European Research Council. Neither the European Union nor the granting authority can be held responsible for them.
  The TESS data presented in this paper were obtained from the Mikulski Archive for Space Telescopes (MAST) at the Space Telescope Science Institute (STScI), which is operated by the Association of Universities for Research in Astronomy, Inc., under NASA contract NAS5-26555. Support to MAST for these data is provided by the NASA Office of Space Science via grant NAG5-7584 and by other grants and contracts. Funding for the TESS mission was provided by the NASA Explorer Program.
\end{acknowledgements}

\bibliography{references}

\begin{thebibliography}{84}
\expandafter\ifx\csname natexlab\endcsname\relax\def\natexlab#1{#1}\fi

\bibitem[{Aerts \& De~Cat(2003)}]{aerts-cep-2003}
Aerts, C. \& De~Cat, P. 2003, Space Science Reviews, 105, 453

\bibitem[{Aerts {et~al.}(2019)Aerts, Mathis, \& Rogers}]{aerts-angular-2019}
Aerts, C., Mathis, S., \& Rogers, T.~M. 2019, ARA\&A, 57, 35

\bibitem[{Aerts {et~al.}(2018)Aerts, Molenberghs, Michielsen, Pedersen,
  Bj{\"o}rklund, Johnston, Mombarg, Bowman, Buysschaert, P{\'a}pics, Sekaran,
  Sundqvist, Tkachenko, Truyaert, Reeth, \& Vermeyen}]{aerts-forward-2018}
Aerts, C., Molenberghs, G., Michielsen, M., {et~al.} 2018, ApJS, 237, 15

\bibitem[{Aerts {et~al.}(2003)Aerts, Thoul, Daszy{\'n}ska, Scuflaire, Waelkens,
  Dupret, Niemczura, \& Noels}]{aerts-asteroseismology-2003}
Aerts, C., Thoul, A., Daszy{\'n}ska, J., {et~al.} 2003, Science, 300, 1926

\bibitem[{Aerts \& Tkachenko(2024)}]{aerts-asteroseismic-2024}
Aerts, C. \& Tkachenko, A. 2024, A\&A, 692, R1

\bibitem[{Belkacem {et~al.}(2015)Belkacem, Marques, Goupil, Mosser, Sonoi,
  Ouazzani, Dupret, Mathis, \& Grosjean}]{belkacem-angular-2015}
Belkacem, K., Marques, J.~P., Goupil, M.~J., {et~al.} 2015, A\&A, 579, A31

\bibitem[{B{\'e}trisey {et~al.}(2023)B{\'e}trisey, Eggenberger, Buldgen,
  Benomar, \& Bazot}]{betrisey-testing-2023}
B{\'e}trisey, J., Eggenberger, P., Buldgen, G., Benomar, O., \& Bazot, M. 2023,
  A\&A, 673, L11

\bibitem[{Bharati~Das {et~al.}(2024)Bharati~Das, Einramhof, \&
  Bugnet}]{bharatidas-unveiling-2024}
Bharati~Das, S., Einramhof, L., \& Bugnet, L. 2024, A\&A, 690, A217

\bibitem[{Bhattacharya {et~al.}(2024)Bhattacharya, Das, Bugnet, Panda, \&
  Hanasoge}]{bhattacharya-detectability-2024}
Bhattacharya, S., Das, S.~B., Bugnet, L., Panda, S., \& Hanasoge, S.~M. 2024,
  ApJ, 970, 42

\bibitem[{Bowman(2020)}]{bowman-asteroseismology-2020}
Bowman, D.~M. 2020, Frontiers in Astronomy and Space Sciences, 7

\bibitem[{Bowman \& Michielsen(2021)}]{bowman-systematic-2021}
Bowman, D.~M. \& Michielsen, M. 2021, A\&A, 656, A158

\bibitem[{Briquet {et~al.}(2007)Briquet, Morel, Thoul, Scuflaire, Miglio,
  Montalb{\'a}n, Dupret, \& Aerts}]{briquet-asteroseismic-2007}
Briquet, M., Morel, T., Thoul, A., {et~al.} 2007, MNRAS, 381, 1482

\bibitem[{Buchler {et~al.}(1997)Buchler, Goupil, \& Hansen}]{buchler-role-1997}
Buchler, J.~R., Goupil, M.~J., \& Hansen, C.~J. 1997, A\&A, 321, 159

\bibitem[{Buchler {et~al.}(1995)Buchler, Goupil, \&
  Serre}]{buchler-dynamic-1995}
Buchler, J.~R., Goupil, M.~J., \& Serre, T. 1995, A\&A, 296, 405

\bibitem[{Bugnet(2022)}]{bugnet-magnetic-2022}
Bugnet, L. 2022, A\&A, 667, A68

\bibitem[{Bugnet {et~al.}(2021)Bugnet, Prat, Mathis, Astoul, Augustson,
  Garc{\'i}a, Mathur, Amard, \& Neiner}]{bugnet-magnetic-2021}
Bugnet, L., Prat, V., Mathis, S., {et~al.} 2021, A\&A, 650, A53

\bibitem[{Burssens {et~al.}(2023)Burssens, Bowman, Michielsen,
  {Sim{\'o}n-D{\'i}az}, Aerts, Vanlaer, Banyard, Nardetto, Townsend, Handler,
  Mombarg, Vanderspek, \& Ricker}]{burssens-calibration-2023}
Burssens, S., Bowman, D.~M., Michielsen, M., {et~al.} 2023, Nat Astron, 7, 913

\bibitem[{Cantiello {et~al.}(2014)Cantiello, Mankovich, Bildsten,
  {Christensen-Dalsgaard}, \& Paxton}]{cantiello-angular-2014}
Cantiello, M., Mankovich, C., Bildsten, L., {Christensen-Dalsgaard}, J., \&
  Paxton, B. 2014, ApJ, 788, 93

\bibitem[{Ceillier {et~al.}(2013)Ceillier, Eggenberger, Garc{\'i}a, \&
  Mathis}]{ceillier-understanding-2013}
Ceillier, T., Eggenberger, P., Garc{\'i}a, R.~A., \& Mathis, S. 2013, A\&A,
  555, A54

\bibitem[{Chandrasekhar(1933)}]{chandrasekhar-equilibrium-1933}
Chandrasekhar, S. 1933, MNRAS, 93, 390

\bibitem[{{Christensen-Dalsgaard} {et~al.}(1990){Christensen-Dalsgaard}, Schou,
  \& Thompson}]{christensen-dalsgaard-comparison-1990}
{Christensen-Dalsgaard}, J., Schou, J., \& Thompson, M.~J. 1990, MNRAS, 242,
  353

\bibitem[{Deheuvels {et~al.}(2014)Deheuvels, Do{\u g}an, Goupil, Appourchaux,
  Benomar, Bruntt, Campante, Casagrande, Ceillier, Davies, De~Cat, Fu,
  Garc{\'i}a, Lobel, Mosser, Reese, Regulo, Schou, Stahn, Thygesen, Yang,
  Chaplin, {Christensen-Dalsgaard}, Eggenberger, Gizon, Mathis,
  {Molenda-{\.Z}akowicz}, \& Pinsonneault}]{deheuvels-seismic-2014}
Deheuvels, S., Do{\u g}an, G., Goupil, M.~J., {et~al.} 2014, A\&A, 564, A27

\bibitem[{Deheuvels {et~al.}(2012)Deheuvels, Garc{\'i}a, Chaplin, Basu, Antia,
  Appourchaux, Benomar, Davies, Elsworth, Gizon, Goupil, Reese, Regulo, Schou,
  Stahn, Casagrande, {Christensen-Dalsgaard}, Fischer, Hekker, Kjeldsen,
  Mathur, Mosser, Pinsonneault, Valenti, Christiansen, Kinemuchi, \&
  Mullally}]{deheuvels-seismic-2012}
Deheuvels, S., Garc{\'i}a, R.~A., Chaplin, W.~J., {et~al.} 2012, ApJ, 756, 19

\bibitem[{Deheuvels {et~al.}(2023)Deheuvels, Li, Ballot, \&
  Ligni{\`e}res}]{deheuvels-strong-2023}
Deheuvels, S., Li, G., Ballot, J., \& Ligni{\`e}res, F. 2023, A\&A, 670, L16

\bibitem[{Di~Mauro {et~al.}(2016)Di~Mauro, Ventura, Cardini, Stello,
  {Christensen-Dalsgaard}, Dziembowski, Patern{\`o}, Beck, Bloemen, Davies,
  Smedt, Elsworth, Garci{\textbackslash}'a, Hekker, Mosser, \&
  Tkachenko}]{dimauro-internal-2016}
Di~Mauro, M.~P., Ventura, R., Cardini, D., {et~al.} 2016, ApJ, 817, 65

\bibitem[{Di~Mauro {et~al.}(2018)Di~Mauro, Ventura, Corsaro, \&
  Moura}]{dimauro-rotational-2018}
Di~Mauro, M.~P., Ventura, R., Corsaro, E., \& Moura, B. L.~D. 2018, ApJ, 862, 9

\bibitem[{Dupret {et~al.}(2004)Dupret, Thoul, Scuflaire,
  {Daszy{\'n}ska-Daszkiewicz}, Aerts, Bourge, Waelkens, \&
  Noels}]{dupret-asteroseismology-2004}
Dupret, M.-A., Thoul, A., Scuflaire, R., {et~al.} 2004, A\&A, 415, 251

\bibitem[{Dziembowski \& Pamyatnykh(2008)}]{dziembowski-two-2008}
Dziembowski, W.~A. \& Pamyatnykh, A.~A. 2008, MNRAS, 385, 2061

\bibitem[{Eggenberger {et~al.}(2012)Eggenberger, Montalb{\'a}n, \&
  Miglio}]{eggenberger-angular-2012}
Eggenberger, P., Montalb{\'a}n, J., \& Miglio, A. 2012, A\&A, 544, L4

\bibitem[{Espinosa~Lara \& Rieutord(2013)}]{espinosalara-selfconsistent-2013}
Espinosa~Lara, F. \& Rieutord, M. 2013, A\&A, 552, A35

\bibitem[{Fritzewski {et~al.}(2025)Fritzewski, Vanrespaille, Aerts, Guo, Hey,
  \& Ridder}]{fritzewski-mode-2025}
Fritzewski, D.~J., Vanrespaille, M., Aerts, C., {et~al.} 2025, A\&A, 698, A253

\bibitem[{Fuller {et~al.}(2014)Fuller, Lecoanet, Cantiello, \&
  Brown}]{fuller-angular-2014}
Fuller, J., Lecoanet, D., Cantiello, M., \& Brown, B. 2014, ApJ, 796, 17

\bibitem[{Fuller {et~al.}(2019)Fuller, Piro, \& Jermyn}]{fuller-slowing-2019}
Fuller, J., Piro, A.~L., \& Jermyn, A.~S. 2019, MNRAS, 485, 3661

\bibitem[{Goldstein \& Townsend(2020)}]{goldstein-contour-2020}
Goldstein, J. \& Townsend, R. H.~D. 2020, ApJ, 899, 116

\bibitem[{Gough \& Thompson(1990)}]{gough-effect-1990}
Gough, D.~O. \& Thompson, M.~J. 1990, MNRAS, 242, 25

\bibitem[{Goupil {et~al.}(2013)Goupil, Mosser, Marques, Ouazzani, Belkacem,
  Lebreton, \& Samadi}]{goupil-seismic-2013}
Goupil, M.~J., Mosser, B., Marques, J.~P., {et~al.} 2013, A\&A, 549, A75

\bibitem[{Guo {et~al.}(2024)Guo, Bedding, Pamyatnykh, Kurtz, Li, Gautam,
  Murphy, \& Aerts}]{guo-oscillation-2024}
Guo, Z., Bedding, T.~R., Pamyatnykh, A.~A., {et~al.} 2024, MNRAS, 535, 2927

\bibitem[{Hansen(2001)}]{hansen-lcurve-2001}
Hansen, P.~C. 2001, in Computational {{Inverse Problems}} in
  {{Electrocardiology}}, 119--142

\bibitem[{Hatta {et~al.}(2022)Hatta, Sekii, Benomar, \&
  Takata}]{hatta-bayesian-2022}
Hatta, Y., Sekii, T., Benomar, O., \& Takata, M. 2022, ApJ, 927, 40

\bibitem[{Hatta {et~al.}(2019)Hatta, Sekii, Takata, \&
  Kurtz}]{hatta-twodimensional-2019}
Hatta, Y., Sekii, T., Takata, M., \& Kurtz, D.~W. 2019, ApJ, 871, 135

\bibitem[{Jermyn {et~al.}(2023)Jermyn, Bauer, Schwab, Farmer, Ball, Bellinger,
  Dotter, Joyce, Marchant, Mombarg, Wolf, Wong, Cinquegrana, Farrell, Smolec,
  Thoul, Cantiello, Herwig, Toloza, Bildsten, Townsend, \&
  Timmes}]{jermyn-modules-2023}
Jermyn, A.~S., Bauer, E.~B., Schwab, J., {et~al.} 2023, ApJS, 265, 15

\bibitem[{Kurtz {et~al.}(2014)Kurtz, Saio, Takata, Shibahashi, Murphy, \&
  Sekii}]{kurtz-asteroseismic-2014}
Kurtz, D.~W., Saio, H., Takata, M., {et~al.} 2014, MNRAS, 444, 102

\bibitem[{Lee \& Baraffe(1995)}]{lee-pulsational-1995}
Lee, U. \& Baraffe, I. 1995, A\&A, 301, 419

\bibitem[{Li {et~al.}(2022)Li, Deheuvels, Ballot, \&
  Ligni{\`e}res}]{li-magnetic-2022}
Li, G., Deheuvels, S., Ballot, J., \& Ligni{\`e}res, F. 2022, Nature, 610, 43

\bibitem[{Loi(2021)}]{loi-topology-2021}
Loi, S.~T. 2021, MNRAS, 504, 3711

\bibitem[{Maeder(2009)}]{maeder-physics-2009}
Maeder, A. 2009, Physics, {{Formation}} and {{Evolution}} of {{Rotating
  Stars}}, ed. G.~B{\"o}rner, A.~Burkert, W.~B. Burton, M.~A. Dopita,
  A.~Eckart, T.~Encrenaz, E.~K. Grebel, B.~Leibundgut, A.~Maeder, \&
  V.~Trimble, Astronomy and {{Astrophysics Library}} (Berlin, Heidelberg:
  Springer)

\bibitem[{Mahalanobis(1936)}]{mahalanobis-generalized-1936}
Mahalanobis, P.~C. 1936, Proceedings of the National Institute of Science of
  India, 2, 49

\bibitem[{Marques {et~al.}(2013)Marques, Goupil, Lebreton, Talon, Palacios,
  Belkacem, Ouazzani, Mosser, Moya, Morel, Pichon, Mathis, Zahn,
  {Turck-Chi{\`e}ze}, \& Nghiem}]{marques-seismic-2013}
Marques, J.~P., Goupil, M.~J., Lebreton, Y., {et~al.} 2013, A\&A, 549, A74

\bibitem[{Mathis \& Bugnet(2023)}]{mathis-asymmetries-2023}
Mathis, S. \& Bugnet, L. 2023, A\&A, 676, L9

\bibitem[{Mathis {et~al.}(2021)Mathis, Bugnet, Prat, Augustson, Mathur, \&
  Garcia}]{mathis-probing-2021}
Mathis, S., Bugnet, L., Prat, V., {et~al.} 2021, A\&A, 647, A122

\bibitem[{Mathis \& Zahn(2005)}]{mathis-transport-2005}
Mathis, S. \& Zahn, J.-P. 2005, A\&A, 440, 653

\bibitem[{Mestel \& Weiss(1987)}]{mestel-magnetic-1987}
Mestel, L. \& Weiss, N.~O. 1987, MNRAS, 226, 123

\bibitem[{Mombarg {et~al.}(2023)Mombarg, Rieutord, \&
  Lara}]{mombarg-first-2023}
Mombarg, J. S.~G., Rieutord, M., \& Lara, F.~E. 2023, A\&A, 677, L5

\bibitem[{Moyano {et~al.}(2022)Moyano, Eggenberger, Meynet, Gehan, Mosser,
  Buldgen, \& Salmon}]{moyano-asteroseismology-2022}
Moyano, F.~D., Eggenberger, P., Meynet, G., {et~al.} 2022, A\&A, 663, A180

\bibitem[{Ouazzani {et~al.}(2012)Ouazzani, Dupret, \&
  Reese}]{ouazzani-pulsations-2012}
Ouazzani, R.-M., Dupret, M.-A., \& Reese, D.~R. 2012, A\&A, 547, A75

\bibitem[{Ouazzani {et~al.}(2019)Ouazzani, Marques, Goupil, Christophe, Antoci,
  Salmon, \& Ballot}]{ouazzani-doradus-2019}
Ouazzani, R.-M., Marques, J.~P., Goupil, M.-J., {et~al.} 2019, A\&A, 626, A121

\bibitem[{Pamyatnykh {et~al.}(2004)Pamyatnykh, Handler, \&
  Dziembowski}]{pamyatnykh-asteroseismology-2004}
Pamyatnykh, A.~A., Handler, G., \& Dziembowski, W.~A. 2004, MNRAS, 350, 1022

\bibitem[{Paxton {et~al.}(2011)Paxton, Bildsten, Dotter, Herwig, Lesaffre, \&
  Timmes}]{paxton-modules-2011}
Paxton, B., Bildsten, L., Dotter, A., {et~al.} 2011, ApJS, 192, 3

\bibitem[{Paxton {et~al.}(2013)Paxton, Cantiello, Arras, Bildsten, Brown,
  Dotter, Mankovich, Montgomery, Stello, Timmes, \&
  Townsend}]{paxton-modules-2013}
Paxton, B., Cantiello, M., Arras, P., {et~al.} 2013, ApJS, 208, 4

\bibitem[{Paxton {et~al.}(2015)Paxton, Marchant, Schwab, Bauer, Bildsten,
  Cantiello, Dessart, Farmer, Hu, Langer, Townsend, Townsley, \&
  Timmes}]{paxton-modules-2015}
Paxton, B., Marchant, P., Schwab, J., {et~al.} 2015, ApJS, 220, 15

\bibitem[{Paxton {et~al.}(2018)Paxton, Schwab, Bauer, Bildsten, Blinnikov,
  Duffell, Farmer, Goldberg, Marchant, Sorokina, Thoul, Townsend, \&
  Timmes}]{paxton-modules-2018}
Paxton, B., Schwab, J., Bauer, E.~B., {et~al.} 2018, ApJS, 234, 34

\bibitem[{Paxton {et~al.}(2019)Paxton, Smolec, Schwab, Gautschy, Bildsten,
  Cantiello, Dotter, Farmer, Goldberg, Jermyn, Kanbur, Marchant, Thoul,
  Townsend, Wolf, Zhang, \& Timmes}]{paxton-modules-2019}
Paxton, B., Smolec, R., Schwab, J., {et~al.} 2019, ApJS, 243, 10

\bibitem[{Pijpers \& Thompson(1994)}]{pijpers-sola-1994}
Pijpers, F.~P. \& Thompson, M.~J. 1994, A\&A, 281, 231

\bibitem[{Reese {et~al.}(2006)Reese, Ligni{\`e}res, \&
  Rieutord}]{reese-acoustic-2006}
Reese, D., Ligni{\`e}res, F., \& Rieutord, M. 2006, A\&A, 455, 621

\bibitem[{Ricker {et~al.}(2014)Ricker, Winn, Vanderspek, Latham, Bakos, Bean,
  {Berta-Thompson}, Brown, Buchhave, Butler, Butler, Chaplin, Charbonneau,
  {Christensen-Dalsgaard}, Clampin, Deming, Doty, Lee, Dressing, Dunham, Endl,
  Fressin, Ge, Henning, Holman, Howard, Ida, Jenkins, Jernigan, Johnson,
  Kaltenegger, Kawai, Kjeldsen, Laughlin, Levine, Lin, Lissauer, MacQueen,
  Marcy, McCullough, Morton, Narita, Paegert, Palle, Pepe, Pepper, Quirrenbach,
  Rinehart, Sasselov, Sato, Seager, Sozzetti, Stassun, Sullivan, Szentgyorgyi,
  Torres, Udry, \& Villasenor}]{ricker-transiting-2014}
Ricker, G.~R., Winn, J.~N., Vanderspek, R., {et~al.} 2014, JATIS, 1, 014003

\bibitem[{Rogers(2015)}]{rogers-differential-2015}
Rogers, T.~M. 2015, ApJL, 815, L30

\bibitem[{Rogers {et~al.}(2013)Rogers, Lin, McElwaine, \&
  Lau}]{rogers-internal-2013}
Rogers, T.~M., Lin, D. N.~C., McElwaine, J.~N., \& Lau, H. H.~B. 2013, ApJ,
  772, 21

\bibitem[{Saio(1981)}]{saio-rotational-1981}
Saio, H. 1981, ApJ, 244, 299

\bibitem[{Salmon {et~al.}(2022)Salmon, Moyano, Eggenberger, Haemmerl{\'e}, \&
  Buldgen}]{salmon-backtracing-2022}
Salmon, S. J. a.~J., Moyano, F.~D., Eggenberger, P., Haemmerl{\'e}, L., \&
  Buldgen, G. 2022, A\&A, 664, L1

\bibitem[{Schatzman(1993)}]{schatzman-transport-1993}
Schatzman, E. 1993, Astronomy and Astrophysics, 279, 431

\bibitem[{Soufi {et~al.}(1998)Soufi, Goupil, \&
  Dziembowski}]{soufi-effects-1998}
Soufi, F., Goupil, M.~J., \& Dziembowski, W.~A. 1998, A\&A, 334, 911

\bibitem[{Spruit(2002)}]{spruit-dynamo-2002}
Spruit, H.~C. 2002, A\&A, 381, 923

\bibitem[{Stankov \& Handler(2005)}]{stankov-catalog-2005}
Stankov, A. \& Handler, G. 2005, ApJS, 158, 193

\bibitem[{Sterken \& Jerzykiewicz(1993)}]{sterken-beta-1993}
Sterken, C. \& Jerzykiewicz, M. 1993, Space Science Reviews, 62, 95

\bibitem[{Su{\'a}rez {et~al.}(2006)Su{\'a}rez, Goupil, \&
  Morel}]{suarez-effects-2006}
Su{\'a}rez, J.~C., Goupil, M.~J., \& Morel, P. 2006, A\&A, 449, 673

\bibitem[{Su{\'a}rez {et~al.}(2009)Su{\'a}rez, Moya, Amado, {Mart{\'i}n-Ruiz},
  {Rodr{\'i}guez-L{\'o}pez}, \& Garrido}]{suarez-seismology-2009}
Su{\'a}rez, J.~C., Moya, A., Amado, P.~J., {et~al.} 2009, ApJ, 690, 1401

\bibitem[{Takahashi \& Langer(2021)}]{takahashi-modeling-2021}
Takahashi, K. \& Langer, N. 2021, A\&A, 646, A19

\bibitem[{Talon \& Charbonnel(2005)}]{talon-hydrodynamical-2005}
Talon, S. \& Charbonnel, C. 2005, A\&A, 440, 981

\bibitem[{Talon {et~al.}(2002)Talon, Kumar, \& Zahn}]{talon-angular-2002}
Talon, S., Kumar, P., \& Zahn, J.-P. 2002, ApJ, 574, L175

\bibitem[{Townsend {et~al.}(2018)Townsend, Goldstein, \&
  Zweibel}]{townsend-angular-2018}
Townsend, R. H.~D., Goldstein, J., \& Zweibel, E. 2018, MNRAS, 475, 879

\bibitem[{Townsend \& Teitler(2013)}]{townsend-gyre-2013}
Townsend, R. H.~D. \& Teitler, S.~A. 2013, MNRAS, 435, 3406

\bibitem[{Triana {et~al.}(2017)Triana, Corsaro, Ridder, Bonanno, Hern{\'a}ndez,
  \& Garc{\'i}a}]{triana-internal-2017}
Triana, S.~A., Corsaro, E., Ridder, J.~D., {et~al.} 2017, A\&A, 602, A62

\bibitem[{Triana {et~al.}(2015)Triana, Moravveji, P{\'a}pics, Aerts, Kawaler,
  \& {Christensen-Dalsgaard}}]{triana-internal-2015}
Triana, S.~A., Moravveji, E., P{\'a}pics, P.~I., {et~al.} 2015, ApJ, 810, 16

\bibitem[{Zahn {et~al.}(1997)Zahn, Talon, \& Matias}]{zahn-angular-1997}
Zahn, J.~P., Talon, S., \& Matias, J. 1997, Astronomy \& Astrophysics, 322, 320

\end{thebibliography}

\begin{appendix}
  \section{New mode frequencies}
  \label{sec:frequencies}
  The frequencies of the modelled multiplets derived from the combined cycle 2 and cycles 4 and 5 TESS observations can be found in Table~\ref{table:frequencies}.
  We use the same names as \citet{burssens-calibration-2023} for these multiplets.
  While more frequencies can be found in this region, for example at \SI{3.45}{\per\day} and \SI{4.03}{\per\day}, which do not fit in the existing pattern and hint at a third multiplet, we consider interpreting these frequencies beyond the scope of this work, which focuses on the rotation inversion.

  \begin{table}
    \caption{Frequencies and spacings of the detected multiplets using the cycle 2 and cycles 4 and 5 TESS data. The frequencies used as zonal modes are indicated in bold.}
    \label{table:frequencies}
    \centering
    \begin{tabular}{l l l}
      \hline
      \hline
      & Frequency [\SI{}{\per\day}] & Spacing [\SI{}{\per\day}] \\
      \hline
      quint1a & 3.445693(7)  & 0.195456(8) \\
              & 3.641149(5)  & 0.195178(14) \\
              & \textbf{3.836328(13)} & 0.191513(14) \\
              & 4.027842(7)  & 0.193815(19) \\
              & 4.221657(17) & \\
              \hline
      quint1b & 3.625608(8)  & 0.193809(13) \\
              & 3.819418(10) & 0.195235(18)\tablefootmark{a} \\
              & \textbf{4.014653(18)} & \textit{not detected}\\
              & 4.209888(15) & \\
              \hline
      quad1 & 6.3509259(4)  & 0.17407779(10) \\
            & 6.5250037(3)  & 0.17180657(8) \\
            & \textbf{6.6968102(10)} & 0.17017918(10) \\
            & 6.8669894(8)  & \\
            \hline
      trip1 & 7.01910(2)   & 0.17015(2)  \\
            & \textbf{7.189260(2)}  & 0.170295(2) \\
            & 7.3595565(2) & \\
            \hline
      trip2 & 7.30484(2)  & 0.17116(2) \\
            & 7.476004(3) & 0.171048(5) \\
            & \textbf{7.647053(4)} & \\
      \hline
    \end{tabular}
    \tablefoot{
    \tablefoottext{a}{As the next mode in the multiplet is not detected, the spacing is given by dividing the spacing with the $m + 2$ mode by two.}
  }
  \end{table}

  % \lfig{HD192575_final_mahalanobis_parameter_correlations}{Posterior distribution (panels on the diagonal) of and correlations (panels below the diagonal) between the parameters fitted in the forward modelling. The color indicates how well the model fits the data as given by the MD merit function. The best model is indicated by a white star. The panel in top right corner shows the $3\sigma$ and $1\sigma$ spectroscopic error boxes. The black crosses indicate the best five models other than the best model indicated by the white star.}

  \section{Rotation profiles}
  \label{sec:all-rotation-profiles}

  The rotation profiles for all of the models obtained from the forward models can be found in Figures \ref{fig:rotation-profiles/spline_rls_rotation_profiles_increased_npg} to \ref{fig:rotation-profiles/spline_rls_rotation_profiles_final_set_6}.

  \sfig{rotation-profiles/spline_rls_rotation_profiles_original}{Rotation profiles obtained from the inversion method described in Sect.~\ref{sec:inversion} for the ``original'' model set. The profiles shown have the lowest level of differential rotation while still fitting the observed splittings.}

  \sfig{rotation-profiles/spline_rls_rotation_profiles_increased_npg}{Rotation profiles obtained from the inversion method described in Sect.~\ref{sec:inversion} for the ``new frequencies'' model set. The profiles shown have the lowest level of differential rotation while still fitting the observed splittings.}
  
  \sfig{rotation-profiles/spline_rls_rotation_profiles_final_set_1}{Rotation profiles obtained from the inversion method described in Sect.~\ref{sec:inversion} for model set 1. The profiles shown have the lowest level of differential rotation while still fitting the observed splittings. The blue profiles have asymmetries for the ``quad1'' multiplet which are similar to the observed asymmetries (see Sect.~\ref{sec:observed-asymmetries} for details).}

  \sfig{rotation-profiles/spline_rls_rotation_profiles_final_set_2}{Rotation profiles obtained from the inversion method described in Sect.~\ref{sec:inversion} for model set 2. The profiles shown have the lowest level of differential rotation while still fitting the observed splittings.}

  \sfig{rotation-profiles/spline_rls_rotation_profiles_final_set_3}{Rotation profiles obtained from the inversion method described in Sect.~\ref{sec:inversion} for model set 3. The profiles shown have the lowest level of differential rotation while still fitting the observed splittings. The blue profiles have asymmetries for the ``quad1'' multiplet which are similar to the observed asymmetries (see Sect.~\ref{sec:observed-asymmetries} for details).}
  
  \sfig{rotation-profiles/spline_rls_rotation_profiles_final_set_4}{Rotation profiles obtained from the inversion method described in Sect.~\ref{sec:inversion} for model set 4. The profiles shown have the lowest level of differential rotation while still fitting the observed splittings.}
  
  \sfig{rotation-profiles/spline_rls_rotation_profiles_final_set_5}{Rotation profiles obtained from the inversion method described in Sect.~\ref{sec:inversion} for model set 5. The profiles shown have the lowest level of differential rotation while still fitting the observed splittings. The blue and orange profiles have asymmetries for the ``quad1'' and ``quint1a'' multiplet respectively which are similar to the observed asymmetries (see Sect.~\ref{sec:observed-asymmetries} for details).}

  \sfig{rotation-profiles/spline_rls_rotation_profiles_final_set_6}{Rotation profiles obtained from the inversion method described in Sect.~\ref{sec:inversion} for model set 6. The profiles shown have the lowest level of differential rotation while still fitting the observed splittings. The blue profiles have asymmetries for the ``quad1'' multiplet which are similar to the observed asymmetries (see Sect.~\ref{sec:observed-asymmetries} for details).}

  \section{Solving the quadratic eigenvalue problem}
  \label{sec:quadratic-eigenvalue-problem}

  The main oscillation equation relevant for this paper is obtained by perturbing the equation of motion in the presence of a velocity field due to the rotation:
  \begin{multline}
    - \rho (\omega + m\Omega)^2 \bm{\xi} + 2i\rho(\omega + m\Omega)\Omega\bm{e}_z \cross \bm{\xi} = \\
    -\bm{\nabla} p' + \rho' \left(\bm{g} + \Omega^2 s \bm{e}_s\right)  + \rho \left(\bm{g}' - \left(\bm{\xi} \cdot \bm{\nabla} \Omega^2\right) s \bm{e}_s\right)\,.
    \label{eqn:the-oscillation-equation}
  \end{multline}
  For ease of notation, we write this equation as
  \begin{equation}
    \left(-\omega^2 + \omega \mathcal{R} + \mathcal{L} + \mathcal{V}\right) \bm{\xi} = 0\,,
    \label{eqn:operator}
  \end{equation}
  where the operator $\mathcal{R}$ contains the terms linear in $\Omega$, $\mathcal{V}$ the terms quadratic in $\Omega$, and $\mathcal{L}$ the terms in the oscillation equation in absence of rotation.
  Given the assumption that the solutions to the rotating case are given by the linear combination
  \begin{equation}
    \bm{\xi}^{\rm rot}_{n} = \sum_{n'}a_{nn'} \bm{\xi}_{n'}\,,
  \end{equation}
  we get the following set of equations by multiplying Eqn.~\eqref{eqn:operator} by $\bm{\xi}_{n}$ from the left:
  \begin{equation}
    \sum_{n'} \left(-\omega^2 \bm{D}_{nn'} + \omega \bm{R}_{nn'} + \bm{L}_{nn'} + \bm{V}_{nn'}\right) a_{n} = 0\,,
  \end{equation}
  for each $n$.
  The matrices $\bm{D}_{nn'}, \bm{R}_{nn'}, \bm{L}_{nn'}$, and $\bm{V}_{nn'}$ are given by the inner products $\left<\bm{\xi}_n, \bm{X}\bm{\xi}_{n'}\right>$ with $\bm{X} = \mathbb{1}, \mathcal{R}, \mathcal{L}$, and $\mathcal{V}$ respectively.
  In the case of a vacuum outer boundary condition, the matrix $\bm{D}$ will be the identity matrix barring numerical errors, but this is not the case for more realistic boundary conditions.
  This can then be expressed as the generalized eigenvalue problem
  \begin{equation}
    \left(-\omega^2 \bm{D} + \omega \bm{R} + \bm{L} + \bm{V}\right) \cdot \bm{a} = 0\,.
  \end{equation}
  Since this eigenvalue problem is quadratic in the eigenvalue (the frequency $\omega$), we first convert this problem to a normal eigenvalue problem.
  Consider that $\bm{a}_1$ is a solution to the above equation with eigenvalue $\omega$.
  We define $\bm{a}_2 = \omega \bm{a}_1$ as a new vector, which leads us to the equation
  \begin{equation}
    \left(-\omega \bm{D} + \bm{R} \right) \cdot \bm{a}_2 + \left(\bm{L} + \bm{V}\right) \cdot \bm{a}_1 = 0\,.
  \end{equation}
  We combine these definition and equation as
  \begin{equation}
    \omega\begin{pmatrix}
      \mathbb{1} & 0 \\
      0 & -\bm{D}
    \end{pmatrix} \cdot \begin{pmatrix} \bm{a}_1 \\  \bm{a}_2 \end{pmatrix}
    +
    \begin{pmatrix}
      0 & -\mathbb{1} \\
      \bm{L} + \bm{V} & \bm{R}
      \end{pmatrix} \cdot \begin{pmatrix} \bm{a}_1 \\ \bm{a}_2 \end{pmatrix} = 0\,,
  \end{equation}
  which is a linear generalized eigenvalue problem, and can be solved with standard methods.

  \section{All remaining asymmetry plots}
  \label{sec:all-asymmetries}

  Figures \ref{fig:rotation-asymmetries/spline_rls_o_c_quad1_a0} to \ref{fig:rotation-asymmetries/spline_rls_o_c_trip2_a2} show the asymmetries for model sets 1 through 6, for the multiplets and components of the multiplets that were not shown in the main text.
  
  \sfig{rotation-asymmetries/spline_rls_o_c_quad1_a0}{Distribution of the theoretical $\mathcal{A}_0$ asymmetries for the ``quad1'' $\ell = 2$ multiplet. Each panel is a different set, indicated in the top left corner of the panel. The orange vertical line indicates the observed asymmetry. Since the multiplet is not complete, only one of the $\mathcal{A}_0$ or the $\mathcal{A}_2$ asymmetry must fit the data. The radial orders associated with these asymmetries can be found in Table~\ref{table:mode-id} under the ``quad1'' multiplet.}
  \sfig{rotation-asymmetries/spline_rls_o_c_quad1_a2}{Same as Fig.~\ref{fig:rotation-asymmetries/spline_rls_o_c_quad1_a0}, but for the $\mathcal{A}_2$ asymmetry.}
  
  \sfig{rotation-asymmetries/spline_rls_o_c_quint1a_a0}{Distribution of the theoretical $\mathcal{A}_0$ asymmetries for the ``quint1a'' $\ell = 2$ multiplet. Each panel is a different set, indicated in the top left corner of the panel. The orange vertical line indicates the observed asymmetry. The radial orders associated with these asymmetries can be found in Table~\ref{table:mode-id} under the ``quinta'' multiplet.}
  \sfig{rotation-asymmetries/spline_rls_o_c_quint1a_a1}{Same as Fig.~\ref{fig:rotation-asymmetries/spline_rls_o_c_quint1a_a0}, but for the $\mathcal{A}_1$ asymmetry.}
  \sfig{rotation-asymmetries/spline_rls_o_c_quint1a_a2}{Same as Fig.~\ref{fig:rotation-asymmetries/spline_rls_o_c_quint1a_a0}, but for the $\mathcal{A}_2$ asymmetry.}
   
  \sfig{rotation-asymmetries/spline_rls_o_c_trip2_a0}{Distribution of the theoretical $\mathcal{A}_0$ asymmetries for the ``trip2'' $\ell = 2$ multiplet. Each panel is a different set, indicated in the top left corner of the panel. The orange vertical line indicates the observed asymmetry. Since the multiplet is not complete, only one of the $\mathcal{A}_0$, $\mathcal{A}_1$, or $\mathcal{A}_2$ asymmetry must fit the data. The radial orders associated with these asymmetries can be found in Table~\ref{table:mode-id} under the ``trip2'' multiplet.}
  \sfig{rotation-asymmetries/spline_rls_o_c_trip2_a1}{Same as Fig.~\ref{fig:rotation-asymmetries/spline_rls_o_c_trip2_a0}, but for the $\mathcal{A}_1$ asymmetry.}
  \sfig{rotation-asymmetries/spline_rls_o_c_trip2_a2}{Same as Fig.~\ref{fig:rotation-asymmetries/spline_rls_o_c_trip2_a0}, but for the $\mathcal{A}_2$ asymmetry.}

\end{appendix}

\end{document}